\documentclass[11pt,a4paper]{article}

\RequirePackage[l2tabu, orthodox]{nag}

\usepackage{jheppub}
\usepackage{amsthm,amssymb,amsmath,epic,float}
\usepackage{rotating,epsfig,indentfirst,array,varioref}
\usepackage{appendix,tikz,mathtools}
\usepackage{setspace}
\usepackage{tikz}
\usetikzlibrary{decorations.pathreplacing}
\usetikzlibrary{calc}
\usepackage{enumitem}
\usepackage{hyphenat}
\usetikzlibrary{positioning}
\usepackage{graphicx}  
\usepackage{adjustbox}

\usepackage{arydshln}

\def\bbt{\bibitem}
\def\be{\begin{equation}}
\def\en{\end{equation}}
\def\ber{\begin{eqnarray}}
\def\enr{\end{eqnarray}}
\def\nmb{ \nonumber\\}
\def\d{\partial}
\def\rbr{\rbrack}
\def\lbr{\lbrack}
\def\rbrc{\rbrace}
\def\lbrc{\lbrace}
\def\ov{\over }
\def\tld{\tilde}

\def\sgm{\sigma}

\def\im{\imath}

\def\dlt{\delta}

\def\bk{{\bf k}}

\def\vw {\vec{w}}
\def\vl{\vec{l}}
\def\vq{\vec{q}}
\def\vt{\vec{t}}
\def\vtau{\vec{\tau}}
\def\vom{\vec{\omega}}
\def\val{\vec{\alpha}}

\def\vk {\vec{k}}

\usepackage{jheppub}
\makeatletter
\def\@fpheader{\vspace{-.1cm}}
\makeatother
                                
\begin{document}

\title{\boldmath  Conformal bootstrap and Mirror symmetry of states in Gepner models.}

\author [a]{Sergej Parkhomenko}

\affiliation[a]{Landau Institute for Theoretical Physics, 142432 Chernogolovka, Russia}


\abstract{We consider two explicit constructions of states in the orbifolds of a product of Minimal $N=(2,2)$ models which are based on twisting by spectral flow, mutual locality and operator algebra requirement. It is shown that these two constructions lead to the Berglund-Hubsh-Krawitz dual orbifold groups which define mirror pairs of isomorphic models. Then we generalize our construction for the orbifolds of Gepner models of superstring compactification and explicitly build IIA/IIB mirror map of the space of states of the superstrings using light-cone gauge.}



\maketitle

\section{Introduction}
\label{sec:intro}

As is well known, 6-dimensional CY manifolds are required to obtain 4-dimensional string models with space-time symmetry \cite{CHSW}. 
An important feature of CY manifolds is mirror symmetry conjecture, which implies the existence of an isomorphism of $\sgm$-models for every mirror pair of CY manifolds and $IIA/IIB$ equivalence of the corresponding superstring compacifications \cite{Witt}-\cite{HKKPT}. This symmetry was predicted by D. Gepner \cite{DG} in the framework of another but equivalent approach to compactification when the degrees of freedom of the $\sgm$-model on the 6-dimensional CY manifold are replaced by an internal $N=(2,2)$ superconformal field theory with central charge $c=9$ \cite{Gep}. Later this symmetry had been discovered in \cite{COGP}, \cite{CLS}, \cite{GrPl}, \cite{LVW}, (see also \cite{LS}).

In paper \cite{GrPl} a construction was developed to show that partition functions of the orbifolds of Fermat-type CY manifolds coincide for each mirror pair. Their idea was based on the observation that orbifold of Minimal model $M_{k}$ over the group $\mathbb{Z}_{k+2}$ gives an equivalent, but mirror reflected model. 

The more refined approach to the mirror symmetry was proposed by Borisov in \cite{B}, \cite{B1}. The author constructed certain vertex algebra for a CY hypersurface in toric manifold, which was endowed with an $N=2$ Virasoro superalgebra action and then proved that vertex algebra of mirror CY hypersurface is isomorphic to the original one.
This paper was an important step towards a rigorous proof of mirror isomorphism for both $N=2$ $\sgm$-models originating from mirror pairs of CY manifolds and for the corresponding $IIA/IIB$ models of superstring compactification.

In the papers \cite{BP},\cite{BBP}, \cite{IIAB} we developed new approach to construct explicitly the space of states for the orbifolds of Gepner models. The approach is based on the mutual locality, operator algebra requirements and spectral flow transformation. The spectral flow automorphisms, which generate the orbifold group $G_{adm}$ and mutual locality requirement were used to construct the states for each twisted sector of the orbifold model.  

In the papers \cite{P}, \cite{IIAB} it was observed that the selection of the mutually local subset of fields from the collection of twisted states is carried out by the dual (mirror) group $G^*_{adm}$. The permutation of $G_{adm}$ and $G^*_{adm}$ is given by the mirror spectral flow  construction of states and transforms the original orbifold into a mirror one and this transformation is an isomorphism of the $N=(2,2)$ superconformal orbifold models \cite{P}. It was implied in \cite{IIAB} that this mirror isomorphism of internal models extends to the IIA/B isomorphism of superstring models, but the details proving this were not presented there.

In this note, we fill this gap by explicitly constructing the $IIA\leftrightarrow IIB$ mirror mapping of states using the light-cone gauge.

The plan of the paper is as follows. In Section~\ref{sec:2} we consider spectral flow construction of states in the compact factor of the superstring following along the papers \cite{BBP}, \cite{P}. This factor is a product $M_{\bk}=\prod_{i=1}^{r}M_{k_{i}}$ of $N=(2,2)$ Minimal models $M_{k_{i}}$ orbifolded by the admissible group $G_{adm}$ which is certain subgroup in the group $\mathbb{Z}_{k_{1}+2}\times...\mathbb{Z}_{k_{r}+2}$ of symmetries of the product model. We start with the spectral flow construction of the primary states in $N=(2,2)$ Minimal model and generalize the construction for the orbifolds of the product models by require the bootstrap axioms satisfaction. In our approach the admissible group condition arises as a requirement that holomorphic and antiholomorphic spin-3/2 currents corresponding to the $(3,0)$ and $(0,3)$ forms of CY manifold are among the fields of the orbifold.

In Section~\ref{sec:3} we discuss mirror pairs of orbifold models, which are determined by a pair of mirror groups ($G_{adm}, G^{*}_{adm}$). We show that, while the twisted sectors are given by the elements the group $G_{adm}$, the set of mutually local fields is given by the elements of dual group $G^{*}_{adm}$. From the other hand, the dual group $G^{*}_{adm}$ appears as the group of twists in the mirror orbifold model, where the group $G_{adm}$ selects the set of mutually local primary fields. 

This picture has already been observed in \cite{GrPl} as a modular invariance condition. However, in this paper we obtain these groups from the bootstrap axioms and spectral flow construction.
Thus, the discussion in section~\ref{sec:3} mainly follows the paper\cite{P}, but here we clarify the connection of mirror symmetry with the spectral flow, the principle of mutual locality and operator algebra.
We proove in particular that mutual locality condition is equivalent to the Berglund-Hubsh-Krawitz dual group definition \cite{BH}, \cite{Kra} (see also \cite{B1}) and the groups $G_{adm}$ and $G^{*}_{adm}$ satisfy this definition. This shows thereby that mirror symmetry follows from bootstrap axioms.

In Section~\ref{sec:4} we consider IIA/IIB Gepner models orbifolds using light-cone gauge and build explicitly the $IIA\leftrightarrow IIB$ mirror mapping of states. We start by writing out the $IIA/IIB$ SUSY algebra generators. Then we derive the equations of generalized GSO projection by requiring that the action of the SUSY algebra is well defined on the physical state space of the string. Rewriting the GSO equations in terms of the groups $G_{adm}, G^{*}_{adm}$ makes these equations mirror symmetric. At the end we extend the mirror mapping construction of the preceding section to explicitly build $IIA\leftrightarrow IIB$ isomorphysm of spaces of states of the string models.

\section{The construction of orbifold states using spectral flow and mutual locality.}
\label{sec:2}

Here, we briefly discuss the explicit construction of states for the orbifolds of the product of $N=(2,2)$ Minimal models developed in \cite{BP}, \cite{BBP}.

We start the discussion considering $N=(2,2)$ superconformal minimal model. The algebra of symmetries of the Minimal model is given by two copies of $N=2$ Virasoro superalgebra with the following commutation relations
\ber
\begin{aligned}
	&\lbr  L_{n},L_{m}\rbr=(n-m)L_{n+m}+{c\ov 12}(n^{3}-n)\dlt_{n+m,0},\\
	&\lbr J_{n},J_{m}\rbr={c\ov 3}n\dlt_{n+m,0},\\
	&\lbr L_{n},J_{m}\rbr=-mJ_{n+m},\\
	&\lbrc G^{+}_{r},G^{-}_{s}\rbrc=L_{r+s}+{r-s\ov 2}J_{r+s}+{c\ov 6}(r^{2}-{1\ov 4})\dlt_{r+s,0},\\
	&\lbr L_{n},G^{\pm}_{r}\rbr=({n\ov 2}-r)G^{\pm}_{r+n},\\
	&\lbr J_{n},G^{\pm}_{r}\rbr=\pm G^{\pm}_{r+n},
	\label{2.N2Vir}
\end{aligned}
\enr
where the central charge is
\ber
c={3k\ov k+2}, \  k=0,1,2,....
\nmb
\enr

The conformal dimensions and charges of the primary 
states $\Phi^{NS}_{l,q}$ in $NS$ sector are
\ber
\Delta_{l,q}=\frac{l(l+2)-q^{2}}{4(k+2)}, \ \ Q_{l,q}=\frac{q}{k+2}.
\label{2.DeltQ}
\enr
In $R$ sector the dimensions and charges of primary states $\Phi^{R}_{l,q}$ are given by
\ber
\Delta^{R}_{l,q}=\frac{l(l+2)-(q-1)^{2}}{4(k+2)}+\frac{1}{8},\quad
\bar{Q}^{R}_{\l,q}=Q^{R}_{l,q}=\frac{q-1}{k+2}+\frac{1}{2}.
\label{2.DeltQR}
\enr
Recall that the chiral primary states \cite{LVW} appear when $q=l$:
\ber
\Phi^{c}_{l}\equiv \Phi^{NS}_{l,l},
\label{2.CPrim}
\enr
while the anti-chiral primary states \cite{LVW} appear when $q=-l$:
\ber
\Phi^{a}_{l}\equiv \Phi^{NS}_{l,-l}.
\label{2.APrim}
\enr

Combaining the holomorphic primary states above with the anti-holomorphic ones we can construct the complete primary fields of the model. In general, the set of (quasi-) local fields of the minimal models have an $A-D-E$ classification, but in this paper we consider only $A$ series.  In this case the primary fields of the model are given by diagonal pairing of holomorphic and anti-holomorphic factors:
\ber
&\Psi^{NS}_{l,q}(z,\bar{z})=\Phi^{NS}_{l,q}(z)\bar{\Phi}^{NS}_{l,q}(\bar{z}),
\\
&\Psi^{R}_{l,q}(z,\bar{z})=\Phi^{R}_{l,q}(z)\bar{\Phi}^{R}_{l,q}(\bar{z}),
\\
& \text{where}\quad l=0,...,k, \quad q=-l,-l+2,...,l.
\label{2.PrimNS}
\enr
Notice that for the $A$ series models there are no chiral-anti-chiral and antichiral-chiral primary fields.

\vskip 10pt
\paragraph{Spectral flow construction of states in $N=(2,2)$ Minimal Model.}


The spectral flow \cite{SS} is given by the one-parametric family of automorphisms
\ber
&\tilde{G}^{\pm}_{r}=U^{-t}G^{\pm}_{r}U^{t}=G^{\pm}_{r\pm t},\\
&\tilde{J}_{n}=U^{-t}J_{n}U^{t}=J_{n}+{c\ov3}t\dlt_{n,0},\\
&\tilde{L}_{n}=U^{-t} L_{n} U^{t}=L_{n}+tJ_{n}+{c\ov 6}t^{2}\dlt_{n,0},
t\in {1\ov 2}+\mathbb{Z}
\label{2.Sflow}
\enr
of the $N=2$ Virasoro superalgebra. 

The spectral flow operator $U^{t}$ can be expressed in terms of a free sacalar field. It is really possible because the second and third commutation relations from (\ref{2.N2Vir}) mean that $J_{n}$ are the Fourier modes of the free bosonic $U(1)$-current $J(z)$, 
which can be written as
\begin{equation}
	J(z)=\im\sqrt{c\ov 3}{\d\phi\ov \d z}.
	\label{2.Jboson}
\end{equation}
The relations (\ref{2.Sflow}) imply then
\begin{equation}
	U^{t}=\exp{(\im t\sqrt{c\ov 3}\phi(z))}.
	\label{2.Uboson}
\end{equation}

For the minimal model $M_{k}$ spectral flow can be used to construct all primary states. Indeed,  as it follows from the singular vector decoupling \cite{FST}, \cite{FeS}, the state 
\ber
V_{l,t}=(UG^{-}_{-{1\ov 2}})^{t}\Phi^{c}_{l}, \quad 0\leq t\leq l.
\label{2.Prim1}
\enr
gives spectral flow realization of the primary state $\Phi^{NS}_{l,q}$, where $q=l-2t$.
For the purposes of the orbifold construction we need to extend this formula. 
Namely, since the corresponding singular vector decouples, the state
\ber
V_{l,t}=(UG^{-}_{-{1\ov 2}})^{t-l-1}U(UG^{-}_{-{1\ov 2}})^{l}\Phi^{c}_{l}, \quad l+1\leq t\leq k+1.
\label{2.Prim2}
\enr
gives in this case the spectral flow realization of the primary state $\Phi^{NS}_{\tld{l},\tld{q}}$, where $\tld{l}=k-l$, $\tld{q}=k+2+l-2t$. 

Notice also the spectral flow the periodicity property \cite{FST}, (see also \cite{FeS})
\ber
U^{k+2}\approx 1.
\label{2.Period}
\enr 
In what follows we will use the notation $V_{l,t}$ instead of $\Phi^{NS}_{l,q}$ to emphasize that we are dealing with the spectral flow realization (\ref{2.Prim1}), (\ref{2.Prim2}) of primary states of Minimal model.

The spectral flow construction of the primary  states in the $R$-sector can be obtained by applying the operator $U^{1\ov 2}$ to the expressions (\ref{2.Prim1}), (\ref{2.Prim2}).

It is worth noting here that, obviously, instead of (\ref{2.Prim1}), (\ref{2.Prim2}) we could propose a construction starting with anti chiral primary fields. This construction will be used in the next section.

\paragraph{The composite of $N=(2,2)$ Minimal Models.}

The orbifolds of composite models 
\ber
M_{\vk}=\prod_{i=1}^{r}M_{k_{i}}
\nmb
\enr 
with the total central charge $9$ are closely related with the
$\sgm$-models on CY manifolds. Following Gepner, we use these orbifolds as a compact sector in the models of superstring compactification.

The $N=(2,2)$ Virasoro superalgebra of the composite model $M_{\vk}$ as well as its orbifold, is defined as the diagonal subalgebra in the tensor product of models:
\ber
L_{orb,n}=\sum_{i}L_{(i),n}, \ J_{orb,n}=\sum_{i}J_{(i),n},
\
G^{\pm}_{orb,r}=\sum_{i}G^{\pm}_{(i),r}.
\label{2.DiagVir}
\enr

The action of this algebra is correctly defined only on the product of $NS$ representations  or on the product of  $R$  representations of minimal models $M_{k_{i}}$.
Therefore, we can form $NS$ or $R$ primary fields  in the product model $M_{\vk}$ by taking only products of primary fields from each minimal model $M_{k_{i}}$  belonging to  the same ($ NS$ or $R$) sectors:

\ber
&\Psi^{NS}_{\vec{l},\vec{q}}(z,\bar{z})=\prod_{i}\Psi^{NS}_{l_{i},q_{i}}(z,\bar{z}),
\
&\Psi^{R}_{\vl,\vq}(z,\bar{z})=\prod_{i}\Psi^{R}_{l_{i},q_{i}}(z,\bar{z}).
\label{2.Prim}
\enr

The descendant fields are generated from these primary fields  by the creation generators of the $N=2$ Virasoro superalgebras of the $M_{k_{i}}$ models. 

\vskip 10pt
\paragraph{The admissible group.}

According to Gepner \cite{Gep}, the orbifolds of composite models are equivalent to $\sgm$-models
on the CY manifolds at certain points of their moduli spaces.


The orbifold is defined by the admissible group \cite{BH}, \cite{Kra}. Recall that 
the composition of the $M_{k_{i}}$ models has a discrete symmetry group
\ber
G_{tot}=\prod_{i=1}^{r}\mathbb{Z}_{k_{i}+2}=\lbrc \prod_{i=1}^{r} \hat{g}_{i}^{w_{i}}, \ w_{i}\in \mathbb{Z}, 
\hat{g}_{i}=\exp{(\im 2\pi J_{(i),0})}\rbrc.
\label{2.Gtot}
\enr

The admissible group is a subgroup of $G_{tot}$ which is defined as follows:
\ber
G_{adm}=\lbrc &\vw=(w^{1},...,w^{r})| \ w^{i}\in\mathbb{Z}_{k_{i}+2},
\
\sum_{i}\frac{w^{i}}{k_{i}+2}\in\mathbb{Z}, \,
\nmb
&\vec{w}= (1,1,1,...,1)\in G_{adm}\rbrc.
\label{2.Gadm}
\enr 
In other words, in addition to the first constraint from this definition we require that the vector $(1,1,1,...,1)\in G_{tot}$ as well as its products be elements of $G_{adm}$. 

As will be seen shortly (see \ref{2.CC1}, \ref{2.CC2}), the constraint
\ber
\sum_{i}\frac{w^{i}}{k_{i}+2}\in\mathbb{Z},
\label{2.CY}
\enr
defining the admissible group, is nothing else but the requirement that the chiral-chiral primaries, charged as $(3,0)$ and $(0,3)$ (which, in addition, are the holomorphic and anti-holomorphic spin-3/2 currents respectively)
are among mutually local fields of the orbifold. Thus, the requirement (\ref{2.CY}) is necessary since these currents must be identified with a nowhere-vanishing holomorphic $(3,0)$ and anti-holomorphic $(0,3)$ forms on the CY hypersurface 
\cite{EOTY}.

\vskip 10pt
\paragraph{Spectral flow construction of primary fields of the orbifold.}

The spectral flow realizations (\ref{2.Prim1}),(\ref{2.Prim2}) are very convenient to build the primary fields in the orbifold model. The construction is given in three steps \cite{BP},\cite{BBP}. 

At the first step, we use the elements $\vw$ of the admissible group (\ref{2.Gadm})
for expanding the state space of the product model $M_{\vk}$ by adding in $NS$ sector the twisted fields of the form
\ber
\Psi^{NS}_{\vl,\vt,\vw}(z,\bar{z})=V_{\vl,\vt+\vw}(z)\bar{V}_{\vl,\vt}(\bar{z}),
\label{2.Primo}
\enr
where
\ber
\bar{V}_{\vl,\vt}(\bar{z})=\prod_{i}\bar{V}_{l_{i},t_{i}}(\bar{z}),\
\bar{V}_{l_{i},t_{i}}(z)=(UG^{-}_{-{1\ov 2}})_{i}^{t_{i}}\bar{\Phi}^{c}_{l_{i}}(z), \ 0\leq t_{i}\leq l_{i},
\label{2.Primo1}
\enr
and
\ber
V_{\vl,\vt+\vw}(z)=\prod_{i=1}^{5}V_{l_{i},t_{i}+w_{i}}(z),
\enr
where
\ber
\begin{aligned}
&V_{l_{i},t_{i}+w_{i}}(z)=\begin{cases}(UG^{-}_{-{1\ov 2}})_{i}^{t_{i}+w_{i}}\Phi^{c}_{l_{i}}(z), 
\qquad \text{if}\quad \ 0\leq t_{i}+w_{i}\leq l_{i}, \\\\
(UG^{-}_{-{1\ov 2}})_{i}^{t_{i}+w_{i}-l_{i}-1}
U_{i}(UG^{-}_{-{1\ov 2}})_{i}^{l_{i}}\Phi^{c}_{l_{i}}(z), 
\quad \text{if}\quad \ l_{i}+1\leq t_{i}+w_{i}\leq k_{i}+1.
 \end{cases}
\label{2.Primo2}
\end{aligned}
\enr

At the second step, we require the mutual locality of the fields obtained above. It gives the equations
\ber
\sum_{i}\frac{w_{1i}(q_{2i}-w_{2i})+
	w_{2i}(q_{1i}-w_{1i})}{k_{i}+2}\in\mathbb{Z},
\label{2.loc}
\enr
where  $\vw_{1}$, $\vw_{2}$ is any pair of twisted sectors from $G_{adm}$ and $\vec{q}_{1,2}=\vl_{1,2}-2\vt_{1,2}$.

The third step of the construction is simple: we generate $R$ sector fields as
\ber
\Psi^{R}_{\vl,\vt,\vw}(z,\bar{z})=\prod_{i}U^{1\ov 2}_{i}\bar{U}^{1\ov 2}_{i}
\Psi^{NS}_{\vl,\vt,\vw}(z,\bar{z}).
\label{2.RPrimo}
\enr

These fields together with the mutually local $NS$ fields constructed before can be considered as a set of primary fields of the orbifolds in the sense that all the other fields of the orbifold are generated by applying to them the creation operators of the $N=(2,2)$ Virasoro superalgebras of $M_{k_{i}}$ models.


The orbifold model satisfy all the requirements of Conformal Bootstrap, as shown in \cite{BBP}. In particular, the OPE is closed on the set of $NS$ sector fields restricted by the locality equations and this entails the OPE closure of the total set of orbifold model fields. 

\vskip 10pt
\paragraph{Chiral-chiral and anti-chiral-chiral primary fields of the orbifold.}

Analysing (\ref{2.Primo})-(\ref{2.Primo2}) one can find what of the primary fields of the orbifold are chiral-chiral or anti-chiral-chiral. Let us recall the construction of these states from \cite{BBP},\cite{P}.

First of all, for the twisted sector $\vw\in G_{adm}$ 
one has to find the vector $\vl=(l_{1},...,l_{5})$ satisfying the equations
\ber
\sum_{i}\tilde{w}_{i}{l_{i}-w_{i}\ov k_{i}+2}\in\mathbb{Z}
\label{2.CC1}
\enr
for any $\tilde{w}\in G_{adm}$.

Then, $(c,c)$ field appears in this sector if for each value of $i$  
\ber
\begin{aligned} 
	&\text{either} \ w_{i}= 0 \mod\ k_{i}+2, \ & \text{or} \ w_{i}=l_{i}+1 \mod\ k_{i}+2,
	\label{2.CC2}
\end{aligned}
\enr
while $(a,c)$ field appears in the sector $\vec{w}$ if for each value of $i$ 
\ber
\begin{aligned}
	&\text{either} \ w_{i}= l_{i} \mod\ k_{i}+2, \ &\text{or} \ w_{i}=k_{i}+1 \mod\ k_{i}+2.
	\label{2.AC2}
\end{aligned}
\enr

\vskip 10pt
\paragraph{Conserved spin-3/2 currents and admissible group.}

Now we show that due to admissible group condition (\ref{2.Gadm}) the orbifold model contains among the mutually local fields the holomorphic and anti-holomorphic spin-3/2 currents which are charged as $(3,0)$ and $(0,3)$.

The holomorphic current is easy to find. It appears in $\vec{w}=(1,1,...,1)$ twisted sector for $\vec{l}=0$. Thus, in according to (\ref{2.CC1}) this current is the charge $(3,0)$ $(c,c)$ primary field and is nothing else but the spectral flow operator 
\ber
\prod_{i}U_{i}(z).
\label{2.30}
\enr

The anti-holomorphic current appears in the twisted sector $\vec{w}=(k_{1}+1,...,k_{r}+1)$ for $\vec{l}=(k_{1},...,k_{r})$. This is charge $(0,3)$ $(c,c)$ primary field and according to (\ref{2.CC2}), (\ref{2.Primo2}) has the form
\ber
\prod_{i}
U_{i}(UG^{-}_{-{1\ov 2}})_{i}^{k_{i}}\Phi^{c}_{l_{i}=k_{i}}(z)\bar{\Phi}^{c}_{l_{i}=k_{i}}(\bar{z}).
\label{2.03}
\enr
Using the commutation relations of $N=2$ Virasoro superalgebra one can write the holomorphic derivative of this field as
\ber
\begin{aligned}
	&L_{orb,-1}\prod_{i}
	U_{i}(UG^{-}_{-\frac{1}{2}} )_{i}^{k_{i}}\Phi^{c}_{l_{i}=k_{i}}(z)\bar{\Phi}^{c}_{l_{i}=k_{i}}(\bar{z})=
	\nmb
	&G^{+}_{orb,-\frac{1}{2}}G^{-}_{orb,-\frac{1}{2}}\prod_{i}
	U_{i}(UG^{-}_{-\frac{1}{2}})_{i}^{k_{i}}\Phi^{c}_{l_{i}=k_{i}}(z)\bar{\Phi}^{c}_{l_{i}=k_{i}}(\bar{z}).
	\nmb
\end{aligned}
\enr
The last expression is zero because of the vector $(G^{-}_{-\frac{1}{2}})_{i}U_{i}(UG^{-}_{-\frac{1}{2}})_{i}^{k_{i}}\Phi^{c}_{l_{i}=k_{i}}(z)$ is singular \cite{FST}, \cite{FeS}. So, this field is an anti-holomorphic current.

The equations (\ref{2.Gadm}), which define the admissible group now follow once we keep the currents (\ref{2.30}), (\ref{2.03}) among the mutually local set of fields.

\section{Mutual locality and Mirror Symmetry in CY orbifolds.}
\label{sec:3}
\vskip 10pt
\paragraph{Mirror spectral flow construction.}

Now we consider mirror spectral flow construction of primary fields of the orbifold model \cite{P}. It starts from anti-chiral primary state in holomorphic sector of (\ref{2.Primo}) and left intact anti-holomorphic factor. Thus, for the anti-holomorphic factor of $\Psi^{NS}_{\vl,\vt,\vw}(z,\bar{z})$ we still have the expression (\ref{2.Primo1}), while for the holomorphic factor we have
\ber
V_{\vl,\vt+\vw}(z)=\prod_{i=1}^{5}V_{l_{i},t_{i}+w_{i}}(z),
\nmb
V_{l_{i},t_{i}+w_{i}}(z)=
(U^{-1}G^{+}_{-{1\ov 2}})_{i}^{l_{i}-t_{i}-w_{i}}\Phi^{a}_{l_{i}}(z), 
\ if \ 0\leq t_{i}+w_{i}\leq l_{i},
\nmb
or
\nmb
V_{l_{i},t_{i}+w_{i}}(z)\equiv
(U^{-1}G^{+}_{-{1\ov 2}})_{i}^{k_{i}+1-t_{i}-w_{i}}
U_{i}^{-1}(U^{-1}G^{+}_{-{1\ov 2}})_{i}^{l_{i}}\Phi^{a}_{l_{i}}(z), 
\nmb
if \ l_{i}+1\leq t_{i}+w_{i}\leq k_{i}+1.
\label{3.Mirror1}
\enr  

Making the involution
\ber
G^{\pm}(z)\rightarrow G^{\mp}(z), \ J(z)\rightarrow -J(z), \ U(z)\rightarrow U^{-1}(z),
\ T(z)\rightarrow T(z).
\label{3.Inv}
\enr
the expressions above transform back into the old form but with different twists
\ber
V_{l_{i},t_{i}+w_{i}}(z)\rightarrow V_{\vl,\vt+\vw^{*}}(z)=\prod_{i=1}^{5}V_{l_{i},t_{i}+w^{*}_{i}}(z),
\nmb
V_{l_{i},t_{i}+w^{*}_{i}}(z)\equiv(UG^{-}_{-{1\ov 2}})_{i}^{t_{i}+w^{*}_{i}}\Phi^{c}_{l_{i}}(z)
=(UG^{-}_{-{1\ov 2}})_{i}^{l_{i}-t_{i}-w_{i}}\Phi^{c}_{l_{i}}(z), 
\ \Rightarrow w^{*}_{i}=l_{i}-2t_{i}-w_{i},
\nmb
or
\nmb
V_{l_{i},t_{i}+w^{*}_{i}}(z)\equiv(UG^{-}_{-{1\ov 2}})_{i}^{t_{i}+w^{*}_{i}-l_{i}-1}
U_{i}(UG^{-}_{-{1\ov 2}})_{i}^{l_{i}}\Phi^{c}_{l_{i}}(z)=
\nmb
(UG^{-}_{-{1\ov 2}})_{i}^{k_{i}+1-t_{i}-w_{i}}
U_{i}(UG^{-}_{-{1\ov 2}})_{i}^{l_{i}}\Phi^{c}_{l_{i}}(z), 
\ \Rightarrow w^{*}_{i}=k_{i}+2+l_{i}-2t_{i}-w_{i}\approx l_{i}-2t_{i}-w_{i}
\nmb
\label{3.Mirror2}
\enr 
(recall that $w^{*}_{i}$ is $mod \ k_{i}+2$ defined, moreover we can rewrite the last expression for $w^{*}_{i}$ as $w^{*}_{i}+w_{i}=\tld{q}_{i}\equiv k_{i}+2+q_{i}$). Recovering the anti-holomorphic factors we obtain
\ber
\Psi^{NS}_{\vl,\vt,\vw^{*}}(z,\bar{z})=V_{\vl,\vt+\vw^{*}}(z)\bar{V}_{\vl,\vt}(\bar{z}).
\label{3.Primdual}
\enr
Thus, the mirror spectral flow construction gives another representation for the fields (\ref{2.Primo}) of the orbifold model $M_{\vk}/G_{adm}$, where the twists are labeled by another set of vectors $\vw^{*}$
\ber
w^{*}_{i}=q_{i}-w_{i}.
\label{3.dualw}
\enr 

\paragraph{Dual model.}
It is easy to see that the new twists form a group because for any pair $\vw^{*}_{1}$ and $\vw^{*}_{2}$, the vector $\vw^{*}_{1}+\vw^{*}_{2}$ arises among the set of $\vw^{*}$. It indeed follows from the OPE since the twists are given by certain products of spectral flow operators so the twisting vectors $\vw^{*}$ are additive w.r.t the OPE. 

Let us denote this group as $G_{adm}^{*}$. It is clear that $G^{*}_{adm}\subset G_{tot}$.

If we rewrite the mutual locality equations (\ref{2.loc}) using the vectors $\vw^{*}$ instead of $\vw$
\ber
\sum_{i}\frac{(q_{1i}-w^{*}_{1i})w^{*}_{2i}+
	(q_{2i}-w^{*}_{2i})w^{*}_{1i}}{k_{i}+2}\in\mathbb{Z},
\label{3.loc}
\enr
the mirror spectral flow construction of the space of states of the orbifold $M_{\vk}/G_{adm}$ will looks like an orbifold $M_{\vk}/G_{adm}^{*}$, where the group $G^{*}_{adm}$ is also admissible since the set of mutually local fields contains the holomorphic and anti-holomorphic spin-3/2 currents (\ref{2.30}), (\ref{2.03}) (although now the first current is charged as $(-3,0)$ due to the involution (\ref{3.Inv})). Moreover, the $N=(2,2)$ superconformal model $M_{\vk}/G^{*}_{adm}$ is isomorphic to the model $M_{\vk}/G_{adm}$.

In addition the following statements take place:


{\bf 1.} The system of mutual locality equations (\ref{2.loc}) is equivalent to the system of equations, which defines Berglund-Hubsh-Krawitz (BHK) mutually dual groups (mutually mirror groups):
\ber
\sum_{i}\frac{w_{i}w^{*}_{i}}{k_{i}+2}\in\mathbb{Z}, \vw\in G_{adm}, \ \vw^{*}\in G_{adm}^{*},
\label{3.BHK}
\enr
so $G_{adm}^{*}$ is Berglund-Hubsh-Krawitz dual group to $G_{adm}$.

{\bf 2.} $(c,c)$ ring of the model  $M_{\vk}/G_{adm}^{*}$ coinsides with the $(a,c)$ ring of the model $M_{\vk}/G_{adm}$ and wise versa.


To prove the statement {\bf 1} we rewite the mutual locality equations (\ref{2.loc}) as the system of equations
\ber
\sum_{i}\frac{w_{1i}w^{*}_{2i}+w_{2i}w^{*}_{1i}}{k_{i}+2}\in\mathbb{Z}, \ for \ each \ pair \ \vw_{1,2}\in G_{adm}.
\label{3.loc1}
\enr
These equations extract for each pair of elements $\vw_{1},\vw_{2}\in G_{adm}$ two sets of vectors $\lbrc\vec{q}_{1}\rbrc$, $\lbrc\vec{q}_{2}\rbrc$ (the vectors $\vq_{1,2}$ enter into $\vw^{*}_{1,2}$ by (\ref{3.dualw})) such that the corresponding fields (\ref{2.Primo}) are mutually local. 

The total set of mutually local fields of the model is obtained by enumerating all pairs $\vw_{1}$, $\vw_{2}$ of twisting vectors from $G_{adm}$ and extracting for each of the pairs two sets of vectors $\lbrc \vec{q}_{1}\rbrc$, $\lbrc \vec{q}_{2}\rbrc$, which correspond to the fields that are mutually local. We thereby obtain the admissible group $G_{adm}^{*}=\lbrc \vec{q}-\vw\rbrc$ as it follows from the mirror spectral flow construction and OPE argument. 

The same is true for the group $G_{adm}$. Namely, for the orbifold model $M_{\vk}/G_{adm}^{*}$ the equations (\ref{3.loc}) identify for each pair of twisting vectors $\vw^{*}_{1},\vw^{*}_{2}\in G_{adm}^{*}$ two sets of vectors $\lbrc\vec{q}_{1}\rbrc$, $\lbrc\vec{q}_{2}\rbrc$ such that the corresponding fields (\ref{3.Primdual}) are mutually local. We obtain the same set of mutually local fields of the orbifold model if one picks over the all pairs from $G_{adm}^{*}$.
In doing so, we iterate over the admissible group $G_{adm}=\lbrc\vec{q}-\vw^{*}\rbrc$. 

Thus, it follows from the two spectral flow constructions that
$U(1)$ charges $\vec{q}$ of the mutually local fields of the orbifold model are given by the pairs $\vw\in G_{adm}, \vw^{*}\in G_{adm}^{*}$ and, by virtue of (\ref{3.dualw}), the field with $U(1)$ charges $\vec{q}$ appears
in the twisted sector $\vw$ ($\vw^{*}$) of the orbifold
$M_{\vk}/G_{adm}$ ($M_{\vk}/G_{adm}^{*}$) if
\ber
\vec{q}=\vw+\vw^{*}.
\label{3.charges}
\enr
Hence, the vectors $\vw_{1,2}\in G_{adm}$, $\vw^{*}_{1,2}\in G_{adm}^{*}$ entering the  equations (\ref{3.loc1}) must be considered as arbitrary taken:
\ber
\sum_{i}\frac{w_{1i}w^{*}_{2i}+w_{2i}w^{*}_{1i}}{k_{i}+2}\in\mathbb{Z}, \ for \ arbitrary \ pairs \ (\vw,\vw^{*})_{1,2}\in G_{adm}\times G_{adm}^{*}.
\label{3.loc2}
\enr
Therefore, this system of equations is equivalent to
\ber
\sum_{i}\frac{w_{i}w^{*}_{i}}{k_{i}+2}\in\mathbb{Z}, \vw\in G_{adm}, \ \vw^{*}\in G_{adm}^{*}.
\label{3.loc3}
\enr

The equations (\ref{3.loc3}) are nothing else but the definition of Berglund-Hubsh- Krawitz dual group for the Fermat type potential $W=X_{1}^{k_{1}+2}+...+X_{r}^{k_{r}+2}$, so the group $G_{adm}^{*}$ is Berglund-Hubsh-Krawitz dual to $G_{adm}$.


Let us now show that equations which define $(c,c)$ field for dual orbifold model $M_{\vk}/G^{*}_{adm}$ are equivalent to the equations which define $(a,c)$ field for the original model and wise versa \cite{P}. 

Because of $t_{i}$ must be zero for the $(c,c)$ field of the dual model, the conditions
(\ref{2.CC2}) for the dual twists take the form
\ber
w^{*}_{i}= 
0 \Leftrightarrow w_{i}=l_{i}, 
\nmb
\text{or}
\ l_{i}+1=w^{*}_{i}\Leftrightarrow w_{i}=k_{i}+1.
\label{3.CC3}
\enr
where we have used expressions for $w^{*}_{i}$ from (\ref{3.dualw}).
Thus, $(c,c)$ equations for the dual model are nothing else but $(a,c)$ equations (\ref{2.AC2}) for the original orbifold.

Similarly for $(a,c)$ field $t_{i}=0$ so that (\ref{2.AC2}) equations for the dual twists take the form
\ber
w^{*}_{i}=l_{i}\Leftrightarrow w_{i}=0,
\nmb
\text{or} \
w^{*}_{i}=k_{i}+1\Leftrightarrow w_{i}=l_{i}+1,
\label{3.AC3}
\enr
where we have used again the expressins for $w^{*}_{i}$ from (\ref{3.dualw}) and have taken into account that $w_{i}$ is $\mod \ k_{i}+2$ defined.
Hence, $(a,c)$ equations for dual model are nothing else but $(c,c)$ equations (\ref{2.CC2}) for the original orbifold. 

Thus, the statement {\bf 2} is proved.

\vskip 10pt
\section{ $IIA\leftrightarrow IIB$ Mirror mapping of states in Gepner models.}
\label{sec:4}

In the light-cone gauge space-time degrees of freedom of type $II$ superstring compactified on the orbifold $M_{\vec{k}}/G_{adm}$ are given by a free-field theory of 2 bosons and 2 Majorana fermions. 

In some appropriate basis the space-time left-moving fields have the OPE
\ber
\d X^{\pm}(z)\d X^{\mp}(0)=-z^{-2}+..., \
\psi^{+}(z)\psi^{-}(0)=z^{-1}+....
\label{4.StOPE}
\enr
Similar OPE's takes place for the right-moving components $\bar{X}^{\pm}(\bar{z})$, $\bar{\psi}^{\pm}(\bar{z})$ of the space-time fields. 

The left-moving $N=2$ Virasoro algebra currents of the space-time factor are given by 
\ber
T_{st}(z)=\d X^{-}\d X^{+}+\frac{1}{2}(\d\psi^{+}\psi^{-}-\psi^{+}\d\psi^{-}),
\nmb
G^{+}_{st}(z)=\psi^{+}(z)\d X^{-}(z), \
G^{-}_{st}(z)=\psi^{-}(z) \d X^{+}(z), 
\nmb
J_{st}(z)=\psi^{+}(z)\psi^{-}(z).
\label{4.StVir}
\enr
The right-moving algebra currents are written similarly.

Thus, on the world sheet we have $N=(2,2)$ superconformal symmetry with total central charge $c_{tot}=12$ which is generated by the diagonal left-moving and right moving $N=2$ Virasoro superalgebras 
\ber
T_{tot}=T_{st}+T_{orb}, \ G^{\pm}_{tot}=G^{\pm}_{st}+G^{\pm}_{orb},
\ J_{tot}=J_{st}+J_{orb},
\label{4.TotVir}
\enr

The algebra action (\ref{4.TotVir}) is correctly defined on the product of only $NS$-representations or on the product of only $R$-representations \cite{Gep}.

\paragraph{Type $IIA/IIB$ space-time SUSY algebra currents.}

It is convenient in what follows to bosonize the fermions
\ber
\psi^{\pm}(z)=\exp{(\pm\im\phi_{0}(z))}, \  
\
\bar{\psi}^{\pm}(\bar{z})=\exp{(\pm\im\bar{\phi}_{0}(\bar{z}))}, 
\nmb
\phi_{0}(z)\phi_{0}(0)=-\log(z)+,...,\
\bar{\phi}_{0}(\bar{z})\bar{\phi}_{0}(0)=-\log(\bar{z})+....
\label{4.FBoson}
\enr

The left-moving SUSY algebra currents for the type $IIA$ case are given by
\ber
{\cal{Q}}^{+}(z)=\exp{[\frac{\im}{2}\phi_{0}-\frac{\im}{2}\sum_{i}\frac{k_{i}}{\sqrt{k_{i}(k_{i}+2)}}\phi_{i}]}(z),
\nmb
{\cal{Q}}^{-}(z)=\exp{[-\frac{\im}{2}\phi_{0}+\frac{\im}{2}\sum_{i}\frac{k_{i}}{\sqrt{k_{i}(k_{i}+2)}}\phi_{i}]}(z),
\nmb
\dot{\cal{Q}}^{+}(z)=
\d X^{-}\exp{[\frac{\im}{2}\phi_{0}-\frac{\im}{2}\sum_{i}\frac{k_{i}\phi_{i}}{\sqrt{k_{i}(k_{i}+2)}}]}(z)+
\nmb
\sum_{i}\sqrt{\frac{k_{i}}{k_{i}+2}}
\hat{G}^{+}_{i}\exp{[-\frac{\im}{2}\phi_{0}-\frac{\im}{2}\sum_{j\neq i}\frac{k_{j}\phi_{j}}{\sqrt{k_{j}(k_{j}+2)}}+\frac{\im}{2}\frac{(k_{i}+4)\phi_{i}}{\sqrt{k_{i}(k_{i}+2)}}]}(z),
\nmb
\dot{\cal{Q}}^{-}(z)=
\d X^{+}\exp{[-\frac{\im}{2}\phi_{0}+\frac{\im}{2}\sum_{i}\frac{k_{i}\phi_{i}}{\sqrt{k_{i}(k_{i}+2)}}]}(z)+
\nmb
\sum_{i}\sqrt{\frac{k_{i}}{k_{i}+2}}
\hat{G}^{-}_{i}\exp{[\frac{\im}{2}\phi_{0}+\frac{\im}{2}\sum_{j\neq i}\frac{k_{j}\phi_{j}}{\sqrt{k_{j}(k_{j}+2)}}-\frac{\im}{2}\frac{(k_{i}+4)\phi_{i}}{\sqrt{k_{i}(k_{i}+2)}}]}(z).
\label{4.SUSYLA}
\enr

The left-moving SUSY algebra currents for the type $IIB$ case are given by
\ber
{\cal{Q}}^{+}(z)=\exp{[\frac{\im}{2}\phi_{0}+\frac{\im}{2}\sum_{i}\frac{k_{i}}{\sqrt{k_{i}(k_{i}+2)}}\phi_{i}]}(z),
\nmb
{\cal{Q}}^{-}(z)=\exp{[-\frac{\im}{2}\phi_{0}-\frac{\im}{2}\sum_{i}\frac{k_{i}}{\sqrt{k_{i}(k_{i}+2)}}\phi_{i}]}(z),
\nmb
\dot{\cal{Q}}^{+}(z)=\d X^{-}\exp{[\frac{\im}{2}\phi_{0}+\frac{\im}{2}\sum_{i}\frac{k_{i}\phi_{i}}{\sqrt{k_{i}(k_{i}+2)}}]}(z)+
\nmb
\sum_{i}\sqrt{\frac{k_{i}}{k_{i}+2}}
\hat{G}^{-}_{i}\exp{[-\frac{\im}{2}\phi_{0}+\frac{\im}{2}\sum_{j\neq i}\frac{k_{j}\phi_{j}}{\sqrt{k_{j}(k_{j}+2)}}-\frac{\im}{2}\frac{(k_{i}+4)\phi_{i}}{\sqrt{k_{i}(k_{i}+2)}}]}(z),
\nmb
\dot{\cal{Q}}^{-}(z)=\d X^{+}\exp{[-\frac{\im}{2}\phi_{0}-\frac{\im}{2}\sum_{i}\frac{k_{i}\phi_{i}}{\sqrt{k_{i}(k_{i}+2)}}]}(z)+
\nmb
\sum_{i}\sqrt{\frac{k_{i}}{k_{i}+2}}
\hat{G}^{+}_{i}\exp{[\frac{\im}{2}\phi_{0}-\frac{\im}{2}\sum_{j\neq i}\frac{k_{j}\phi_{j}}{\sqrt{k_{j}(k_{j}+2)}}+\frac{\im}{2}\frac{(k_{i}+4)\phi_{i}}{\sqrt{k_{i}(k_{i}+2)}}]}(z).
\label{4.SUSYLB}
\nmb
\enr

In the formulas (\ref{4.SUSYLA}), (\ref{4.SUSYLB}) 
we bosonized $U(1)$ factors of the spin $3/2$ currents of $N=2$ Virasoro superalgebra of each Minimal model:
\ber
\bar{G}^{\pm}_{i}(\bar{z})=\exp{(\pm\im\sqrt{\frac{k_{i}+2}{k_{i}}}\bar{\phi}_{i}(\bar{z}))}\hat{\bar{G}}^{\pm}_{i}(\bar{z}).
\nmb
\enr

The right-moving SUSY algebra currents in both cases are given by
\ber
\bar{\cal{Q}}^{+}(\bar{z})=\exp{[\frac{\im}{2}\bar{\phi}_{0}+\frac{\im}{2}\sum_{i}\frac{k_{i}}{\sqrt{k_{i}(k_{i}+2)}}\bar{\phi}_{i}]}(\bar{z}),
\nmb
\bar{\cal{Q}}^{-}(\bar{z})=\exp{[-\frac{\im}{2}\bar{\phi}_{0}-\frac{\im}{2}\sum_{i}\frac{k_{i}}{\sqrt{k_{i}(k_{i}+2)}}\bar{\phi}_{i}]}(\bar{z}),
\nmb
{\dot{\bar{\cal{Q}}}}^{+}(\bar{z})=\bar{\d }\bar{X}^{-}\exp{[\frac{\im}{2}\bar{\phi}_{0}+\frac{\im}{2}\sum_{i}\frac{k_{i}\bar{\phi}_{i}}{\sqrt{k_{i}(k_{i}+2)}}]}(\bar{z})+
\nmb
\sum_{i}\sqrt{\frac{k_{i}}{k_{i}+2}}
\hat{\bar{G}}^{-}_{i}\exp{[-\frac{\im}{2}\bar{\phi}_{0}+\frac{\im}{2}\sum_{j\neq i}\frac{k_{j}\phi_{j}}{\sqrt{k_{j}(k_{j}+2)}}-\frac{\im}{2}\frac{(k_{i}+4)\bar{\phi}_{i}}{\sqrt{k_{i}(k_{i}+2)}}]}(\bar{z}),
\nmb
{\dot{\bar{\cal{Q}}}}^{-}(\bar{z})=\bar{\d} \bar{X}^{+}\exp{[-\frac{\im}{2}\bar{\phi}_{0}-\frac{\im}{2}\sum_{i}\frac{k_{i}\bar{\phi}_{i}}{\sqrt{k_{i}(k_{i}+2)}}]}(\bar{z})+
\nmb
\sum_{i}\sqrt{\frac{k_{i}}{k_{i}+2}}
\hat{\bar{G}}^{+}_{i}\exp{[\frac{\im}{2}\bar{\phi}_{0}-\frac{\im}{2}\sum_{j\neq i}\frac{k_{j}\bar{\phi}_{j}}{\sqrt{k_{j}(k_{j}+2)}}+\frac{\im}{2}\frac{(k_{i}+4)\bar{\phi}_{i}}{\sqrt{k_{i}(k_{i}+2)}}]}(\bar{z}).
\label{4.SUSYR}
\nmb
\enr

 The algebra of the left-moving SUSY $IIA$ currents is given by
\ber
\dot{\cal{Q}}^{+}(z)\dot{\cal{Q}}^{-}(0)=
z^{-3}4-z^{-2}J_{tot}+z^{-1}(T_{tot}-\frac{1}{2} \d J_{tot})+...,
\nmb
\dot{\cal{Q}}^{+}(z){\cal{Q}}^{-}(0)=z^{-1}\d X^{-}(0)+...,\
\dot{\cal{Q}}^{-}(z){\cal{Q}}^{+}(0)=z^{-1}\d X^{+}(0)+...,
\nmb
{\cal{Q}}^{+}(z){\cal{Q}}^{-}(0)=z^{-1}+....
\label{4.SUSYAlg}
\enr

The algebra of $IIB$ currents coincides with (\ref{4.SUSYAlg}) with the only difference in the internal part of $U(1)$ current: $J_{tot}=J_{st}-J_{orb}$.

\paragraph{Space-time SUSY requirement.}

Let us consider complete type $IIB$ $(NS,NS)$ vertices whith some of the descendants omitted
\ber
\Psi^{NS}_{\vl,\vtau,\vom}(z,\bar{z})=
V_{\vec{l},\vtau+\vom}(z)\bar{V}_{\vec{l},\vtau}(\bar{z}),
\nmb
V_{\vec{l},\vtau+\vom}(z)=\exp{[\im t_{0}\phi_{0}]}V_{\vec{l},\vec{t}+\vw}(z),
\
\bar{V}_{\vec{l},\vtau}(\bar{z})=\exp{[\im \bar{t}_{0}\bar{\phi}_{0}]}(\bar{z})\bar{V}_{\vec{l},\vt}(\bar{z}).
\label{4.Psitot}
\enr
Here we introduced the $r+1$ component vectors $\vtau=(t_{0},t_{1},...,t_{r})$, $\vom=(0,w_{1},...,w_{r})$ to include space-time fermionic degrees of freedom.
The vertices above can obviously be written as the product of space-time fermionic contribution and orbifold complete field (\ref{2.Primo})
\ber
\Psi^{NS}_{\vl,\vtau,\vom}(z,\bar{z})=
\exp{[\im t_{0}\phi_{0}]}(z)
\exp{[\im \bar{t}_{0}\bar{\phi}_{0}]}(\bar{z})
\Psi^{NS}_{\vec{l},\vt,\vw}(z,\bar{z}).
\label{4.Psitot1}
\enr

The SUSY requirement can be formulated as a choice of certain representation of the right moving algebra (\ref{4.SUSYR}) on the space of fields (\ref{4.Psitot}). 

Indeed, the first couple of the currents (\ref{4.SUSYR}) from the right-moving sector have conformal dimension ${1\ov 2}$, so they must be considered as fermions on the world-sheet. 
The space-time SUSY requirement in the right-moving sector can be formulated as the condition that the currents $\bar{{\cal{Q}}}^{\pm}(\bar{z})$, $\dot{\bar{{\cal{Q}}}}^{\pm}(\bar{z})$, are in the R sector w.r.t the all physical fields of the string. 

It means in particular that $\bar{{\cal{Q}}}^{+}(\bar{z})\Psi^{NS}_{\vl,\vtau,\vom}(0)$ OPE must have $\bar{z}^{\frac{1}{2}+integer}$ branching:
\ber
\bar{{\cal{Q}}}^{+}(\bar{z})\Psi^{NS}_{\vl,\vtau,\vom}(0)=
\bar{z}^{\frac{1}{2}+integer}\tilde{\Psi}^{NS}_{\vl,\vtau,\vom}(0)+....
\label{4.R}
\enr


The R sector restriction for the space-time SUSY currents $\bar{{\cal{Q}}}^{\pm}(\bar{z})$ is nessesary since zero modes of these currents will act on the fixed $\bar{L}_{tot,0}$-level subspaces giving thereby the space-time SUSY algebra representation on the space of physical states. In other words, GSO-projection means that the physical states of the compactified NSR string is a representation of Green-Schwartz algebra of the fields $\bar{{\cal{Q}}}^{\pm}(\bar{z})$ \cite{GSW}. 

Using (\ref{4.Psitot1}), (\ref{4.SUSYR}) we find
\ber
\bar{{\cal{Q}}}^{+}(\bar{z})\Psi^{NS}_{\vl,\vtau,\vom}(0)=\bar{z}^{\frac{1}{2}(\bar{t}_{0}+\sum_{i}\frac{\bar{q}_{i}}{k_{i}+2})}\Psi^{NS}_{\vl,\vt+\frac{1}{2}\val,\vom-\frac{1}{2}\val}(0)+...=
\bar{z}^{\frac{1}{2}(\bar{t}_{0}+\sum_{i}\frac{\bar{q}_{i}}{k_{i}+2})}V_{\vec{l},\vtau+\vom}(z)\bar{V}_{\vec{l},\vtau+\frac{1}{2}\val}(\bar{z})+...,
\nmb
\enr
where $r+1$ component vector $\val$ is given by  $(1,1,...,1)$, $\bar{q}_{i}=l_{i}-2t_{i}$, $i=1,...,r$.
Hence, the space-time SUSY requirement for the states (\ref{4.Psitot}) is the GSO equation
\ber
\bar{t}_{0}+\sum_{i}\frac{\bar{q}_{i}}{k_{i}+2}=2\bar{N}+1, \ \bar{N}\in\mathbb{Z}.
\label{4.GSOR}
\enr 

In general, the right-moving factor of the string state differs from (\ref{4.Psitot}) by the polynomial:
\ber
P(\bar{\d} \bar{X}^{\pm},\bar{\d}\bar{\phi}_{0},\bar{J}_{i},\bar{T}_{i},\bar{G}^{\pm}_{i}),
\label{4.RPol}
\enr
of the fields $\bar{\d} \bar{X}^{a},\bar{\d}\bar{\phi}_{0},\bar{J}_{i},\bar{T}_{i},\bar{G}^{\pm}_{i}$ and theirs derivatives.

The generalization of (\ref{4.GSOR}) which takes into account descendands (\ref{4.RPol}) has the form
\ber
\bar{t}_{0}+\bar{s}+\sum_{i}\frac{\bar{q}_{i}}{k_{i}+2}=2\bar{N}+1,
\label{4.GSOR1}
\enr
where $\bar{s}\in\mathbb{Z}$ is the charge contribution of internal descendants. It is clear that the second pair of SUSY currents (\ref{4.SUSYR}) is also in $R$ sector provided (\ref{4.GSOR1}) is fulfilled.

We can use (\ref{3.dualw}) to rewrite GSO equation in the form
\ber
\bar{t}_{0}+\bar{s}+\sum_{i}\frac{w^{*}_{i}+w_{i}}{k_{i}+2}=2\bar{N}+1.
\label{4.GSOR2}
\nmb
\enr


Similarly, the space-time SUSY requirement in the left-moving sector is nothing else but R sector restriction for the algebra (\ref{4.SUSYLB}).

To write out the left-moving GSO equation one needs to take into account the twists, labeled by the vectors $\vw$ and the descendants which are given by the polynomials
\ber
P(\d X^{\pm},\d\phi_{0},J_{i},T_{i},G^{\pm}_{i}).
\label{4.LPol}
\enr

It gives the left-moving GSO equation
\ber
t_{0}+s'+\sum_{i}Q_{V_{l_{i},\bar{t}_{i}+w_{i}}}=2N+1,
\label{4.GSOL}
\enr
where $Q_{V_{l_{i},\bar{t}_{i}+w_{i}}}$ is the charge of the primary states $V_{l_{i},\bar{t}_{i}+w_{i}}$ and $s'$ is the charge contribution for the descendant field. Taking into account (\ref{2.Primo2}) the equation above is equivalent to
\ber
t_{0}+\sum_{i}\frac{l_{i}-2t_{i}}{k_{i}+2}-2\sum_{i}\frac{w_{i}}{k_{i}+2}+s=2N+1,
\nmb
\label{4.GSOL1}
\enr
where $s$ differs from $s'$ depending on whether the first or second inequality holds in (\ref{2.Primo2}) for each $i$.

The equation (\ref{4.GSOL1}) can be rewritten in the form
\ber
t_{0}+s+\sum_{i}\frac{w^{*}_{i}-w_{i}}{k_{i}+2}=2N+1.
\enr

So, space-time SUSY requirement for the right and left-moving states in $(NS,NS)$ sector is nothing else but the odd integer $U(1)$ charge projection of D.Gepner \cite{Gep}.

For the $(NS,R)$, $(R,NS)$ and $(R,R)$ sector states the GSO equations follow from the (\ref{4.GSOR1}), (\ref{4.GSOL1}) since, the states from these sectors are generated from the $(NS,NS)$ sector states by the corresponding spectral flows. But these spectral flow operators are the charges of the currents ${\cal{Q}}^{\pm}(z)$ and $\bar{\cal{Q}}^{\pm}(\bar{z})$. These currents are mutually local to each other, so the GSO equations for the states from these sectors are satisfied.

\paragraph{Level matching and BHK dual group equations.}

The physical states of the superstring must satisfy also the level matching condition:
\ber
L_{tot,0}-\bar{L}_{tot,0}=0.
\label{4.Levmatch}
\enr

In $(NS,NS)$ sector the eigenvalue of $L_{tot,0}$ contains the contribution $\Delta_{\vl,\vt+\vw}$ of primary states $V_{\vl,\vt+\vw}$, the contribution of creation operators of $N=2$ Virasoro superalgebras of the minimal models $M_{k_{i}}$ and the contribution of space-time oscillators. The same is true for the eigenvalue of $\bar{L}_{tot,0}$. In order to satisfy (\ref{4.Levmatch}) in $(NS,NS)$ sector the difference  $\Delta_{\vl,\vt+\vw}-\bar{\Delta}_{\vl,\vt}$ must be at least half-integer.

By the direct calculation we find
\ber
\Delta_{l_{i},t_{i}+w_{i}}-\bar{\Delta}_{l_{i},t_{i}}=
\frac{l_{i}(l_{i}+2)-(l_{i}-2(t_{i}+w_{i}))^{2}}{4(k_{i}+2)}-\frac{l_{i}(l_{i}+2)-(l_{i}-2t_{i})^{2}}{4(k_{i}+2)})=
\nmb
\frac{(q_{i}-w_{i})w_{i}}{k_{i}+2}=
\frac{w^{*}_{i}w_{i}}{k_{i}+2}, \ if \ 0\leq t_{i}+w_{i}\leq l_{i},
\nmb
\Delta_{l_{i},t_{i}+w_{i}}-\bar{\Delta}_{l_{i},t_{i}}=
w_{i}+\frac{w^{*}_{i}w_{i}}{k_{i}+2}, \ if \ l_{i}+1\leq t_{i}+w_{i}\leq k_{i}+1.
\nmb
\enr
Therefore the neccessary condition for the level matching is
\ber
\sum_{i}\frac{w^{*}_{i}w_{i}}{k_{i}+2}\in \frac{1}{2}\mathbb{Z}.
\label{4.Levmatch1}
\enr
Due to BHK equations (\ref{3.BHK}) this condition is fulfilled. 

Level matching in the sectors $(NS,R)$, $(R,NS)$, $(R,R)$ is also satisfied, since the states in these sectors are generated by zero modes of the left and right SUSY currents from the $(NS,NS)$ sector states.


\paragraph{$IIA\leftrightarrow IIB$ mirror symmetry mapping.}

Now we can repeat the arguments of the Section 3 applying the mirror spectral flow construction for the compact factor of complete vertices of $IIB$ string:
\ber
\Psi^{NS}_{\vl,\vtau,\vom^{*}}(z,\bar{z})=
V_{\vec{l},\vtau+\vom^{*}}(z)\bar{V}_{\vec{l},\vtau}(\bar{z}),
\nmb
V_{\vec{l},\vtau+\vom^{*}}(z)=\exp{[\im t_{0}\phi_{0}]}V_{\vec{l},\vec{t}+\vw^{*}}(z),
\
\bar{V}_{\vec{l},\vtau}(\bar{z})=\exp{[\im \bar{t}_{0}\bar{\phi}_{0}]}(\bar{z})\bar{V}_{\vec{l},\vt}(\bar{z}),
\label{4.PsitotM}
\enr
where the involution (\ref{3.Inv}) in the compact factor is implied. As a result, as we have seen, the orbifold $M_{\vk}/G_{adm}$ is replaced by the orbifold $M_{\vk}/G^{*}_{adm}$. 

At the same time $IIB$ SUSY algebra of currents (\ref{4.SUSYLB}) transforms into the $IIA$ SUSY algebra of currents (\ref{4.SUSYLA}). Hence, the left-moving and right-moving GSO equations change as
\ber
t_{0}+s+\sum_{i}\frac{w^{*}_{i}-w_{i}}{k_{i}+2}\rightarrow
t_{0}-s+\sum_{i}\frac{w_{i}-w^{*}_{i}}{k_{i}+2}\in 2\mathbb{Z}+1,
\nmb
\bar{t}_{0}+\bar{s}+\sum_{i}\frac{w^{*}_{i}+w_{i}}{k_{i}+2}\rightarrow
\bar{t}_{0}+\bar{s}+\sum_{i}\frac{w_{i}+w^{*}_{i}}{k_{i}+2}\in 2\mathbb{Z}+1.
\enr
It is also clear how the left-moving and the right-moving descendants transform under this involution.

Since the level matching condition (\ref{4.Levmatch1}) is invariant under this mirror transformation we thereby obtain the explicit isomorphysm mapping between the states of the type $IIB$ string, compactified on the orbifold $M_{\vk}/G_{adm}$, and the states of the type $IIA$ string, compactified on the mirror orbifold $M_ {\vk}/G^{*}_{adm}$.

\section{Conclusion}
\label{sec:5}

Using spectral flow twisting and bootstrap axioms we discussed explicit construction of states for the orbifolds of the products $M_{\vec{k}}$ of $N=2$ superconformal Minimal models.  In our approach the orbifold model $M_{\vec{k}}/G_{adm}$ appears simultaineously with its mirror orbifold $M_{\vec{k}}/G^{*}_{adm}$, where $G^{*}_{adm}$, as it was prooved in this paper, is Berglund-Hubsh-Krawitz dual group. It is shown thereby that myrror symmetry of the $\sigma$-models follows from the bootstrap axioms.
 
We generalized the construction to build the physical states of Gepner models of type II superstring compactification in the light-cone gauge and constructed $IIA\leftrightarrow IIB$ mirror isomorphysm mapping of states. 

The approach can be straightforwardly applied for the Heterotic string models. 

It would also be interesting to extend the construction to include the Minimal models of D and E type \cite{ADE}. 

The more difficult problem is to generalize the approach for the type II superstring compactified on toric CY manifolds. The results represented in \cite{B}, \cite{B1} may appear to be helpfull in this concern.

\section*{Acknowledgments} 

Author acknowledges A. A. Belavin and A. V. Litvinov for helpful discussions. 
The work was carried out at Landau Institute for Theoretical Physics in the framework of the state assignment FFWR-2024-0012.


\end{document}